# Electrical and thermal conductivities of reduced graphene oxide/polystyrene composites

Wonjun Park[1, 2, a)], Jiuning Hu[1, 2], Luis A. Jauregui[1, 2], Xiulin Ruan[3, 2], and Yong P. Chen[4, 2, 1, b)]

[1]*School of Electrical and Computer Engineering, Purdue University, West Lafayette, IN 47907, U.S.A.*
[2]*Birck Nanotechnology Center, Purdue University, West Lafayette, IN 47907, U.S.A.*
[3]*School of Mechanical Engineering, Purdue University, West Lafayette, IN 47907, U.S.A.*
[4]*Department of Physics, Purdue University, West Lafayette, IN 47907, U.S.A.*

## Abstract

The author reports an experimental study of electrical and thermal transport in reduced graphene oxide (RGO)/polystyrene (PS) composites. The electrical conductivity (σ) of RGO/PS composites with different RGO concentrations at room temperature shows a percolation behavior with the percolation threshold of ~ 0.25 vol.%. Their temperature-dependent electrical conductivity follows Efros-Shklovskii (ES) variable range hopping (VRH) conduction in the temperature range of 30 to 300 K. The thermal conductivity (κ) of composites is enhanced by ~ 90 % as the concentration is increased from 0 to 10 vol.%. The thermal conductivity of composites approximately linearly increases with increasing temperature from 150 to 300 K. Composites with a higher concentration show a stronger temperature dependence in the thermal conductivity.

[a)] Electronic mail: park249@purdue.edu
[b)] Electronic mail: yongchen@purdue.edu

After the pioneering work of graphene,[1] there have been a lot of efforts to investigate a chemical route to produce exfoliated graphene oxide (GO) and reduced graphene oxide (RGO).[2] Although electrical and thermal properties of RGO are not comparable to those of pristine graphene due to structural defects,[3-5] there are still many advantages in real applications. In general, the chemical approach for producing GO and RGO is suited for a mass production process and a low-cost procedure and especially, it can render us a large variety of different variants of graphene with chemical modifications. For example, RGO has received intense attention as a filler material in a variety of composites or hybrid systems such as batteries,[6] electrodes,[7] photodetectors,[8] or thermal interface materials (TIMs).[9]

Many types of commercial polymer composites have been used for thermal management applications. Especially, electrically and thermally conductive polymer composites are used as heat sinks for device packaging requiring a high thermal conductivity for thermal management and a high electrical conductivity for electromagnetic shielding.[10] In addition, TIMs such thermal greases are used to improve the efficiency of heat conduction at thermal junctions. The thermal conductivity of those TIMs lies between ~ 3 and ~ 8 W/mK.[11] However, these commercial composites are often based on expensive fillers such as metallic/ceramic fillers.

Among many composite systems, graphene-based (e.g. RGO or exfoliated graphene) polymer composites are promising materials for thermal management applications. For instance, we can develop thermally and electrically conductive polymer composites with a cost-efficient way by combining RGO with polystyrene (PS), known as one of the most common and inexpensive polymers for packaging and consumer electronics.

In this work, we made RGO/PS composites fabricated by the chemical reduction method. We characterized electrical ($\sigma$) and thermal ($\kappa$) conductivities of RGO/PS composites in order to explore their feasibility as electrically and thermally conductive composites. This work will provide better understanding of electrical and thermal conductivities of RGO/PS composites at various temperatures and filler concentrations.

Graphite oxide was synthesized from SP-1 graphite powder (average size ~ 75 ± 43 μm, Bay Carbon Inc.) by modified Hummers method.[12] RGO/PS composites were prepared by using a similar method as described in ref. 13, but we used PS with a larger molecular weight ($M_w$). Graphite oxide was treated with phenyl-isocyanate (≥ 98%, Sigma-Aldrich) for 24 hours and it was filtered. The isocyante-treated graphite oxide was dispersed in N,N-dimethylformamide (DMF) (99.8%, Sigma-Aldrich) and ultra-sonicated for 2 hours for exfoliation of graphite oxide into GO. After PS (Scientific Polymer Products Inc., approximate $M_w$ = ~ 400,000) was added to the GO suspension, the GO/PS solution was treated by N,N-dimethylhydrazine (98%, Sigma-Aldrich) at 80 °C for 24 hours for the reduction of GO. After the reduction process, the mixture was polymerized by methanol and the coagulated RGO/PS composites were collected by filtering process. The solid form of composite was dried and ground into fine powder. The composite powder was hot-pressed in the rectangular shape of metal molds at 200 °C for 1 hour.

For a qualitative understanding of structural changes after the reduction process, Raman spectroscopy (using a Horiba Jobin Yvon Xplora confocal Raman microscope) was performed on GO and RGO with a 100 × objective lens (numerical aperture = 0.90) and incident laser (wavelength = 532 nm) power of ~ 1.4 mW. GO flakes were deposited on a $SiO_2$/Si substrate by a spin-coating method. GO was exposed to a N,N-dimethylhydrazine vapor in a sealed petri dish at 80 °C for 24 hours. The thickness of GO on the $SiO_2$/Si substrate was confirmed with an atomic force microscope (AFM) (NT-MDT NTEGRA). Cross-sectional structures of freshly cleaved RGO/PS composites were studied by a scanning electron microscope (SEM) (Hitachi S-4800). For electrical conductivity measurements, composites were cut into rectangular bars and their channel (between electrodes for the voltage probe) dimensions were length × width × thickness = 1 × 1 × 1 $mm^3$ for 0.5, 1, 2.5, 5, 10, and 20 vol.% RGO/PS and 0.15 × 2 × 1 $mm^3$ for 0.25 vol.% RGO/PS (here, the vol.% refers to the concentration of RGO in composites). Cr (20 nm)/Au (180 nm) was deposited as electrodes using e-beam evaporation. Electrical conductivity of composites was measured with a source-meter (Keithley 2400) and a multimeter (HP 34401A) based on a 4-probe measurement. Electrical conductivity at low temperatures was measured inside a variable temperature insert (VTI). Thermal conductivity was measured using the 3ω method. To facilitate such measurements, we polished the surface of composites with alumina particles (50 nm). Composites were spin-coated with a thin layer of polyvinyl alcohol (200 ~ 300 nm) as an insulating layer for conductive composites. Cr (20nm)/Au (180nm) metal lines were deposited by a shadow mask technique. Their width and length were 40 μm and 4 mm, respectively. The 3ω measurement was conducted in a cryogenic stage (Cryo Industries of America Inc.) under vacuum condition (< ~ $10^{-4}$ torr). Lock-in amplifiers (SR 830) were used for collecting 3ω signals after removing the 1ω signal via homemade differential amplifier circuits as described in previous reports.[14, 15] Temperature coefficients of the metal lines were measured after collecting 3ω signals, where the temperature was monitored by a thermometer (cernox) mounted on the top of the sample and a temperature controller (Lakeshore 336). In order to facilitate efficient temperature control, all of samples were mounted on homemade metal chip carriers. As a benchmark, we also measured the thermal conductivity of a borosilicate substrate (pyrex) at the temperature down to 100 K based on the same measurement and data analysis procedure. The thermal conductivity of the borosilicate substrate we measured is in good agreement with the known value (κ ~ 1.14 at room temperature) with 6% uncertainty in the measurement temperature range of 100 to 300 K.[14]

The thickness of freshly exfoliated GO was confirmed with AFM. A representative AFM image of a GO flake is shown in Fig. 1 (a). The height profile along the horizontal line clearly shows two plateaus and the height of each plateau is around 1 nm. The average lateral size of overall GO flakes is measured to be ~ 1 μm and it is similar to the previous result.[13] Fig. 1 (b) shows representative Raman spectra of GO and RGO flakes. The Raman results are also consistent with the previous report.[16] After the reduction of GO, we notice two distinctive features which are related to the chemical reduction of GO. First, I(D)/I(G) of RGO is increased by 6%, which implies that small graphitic domains are formed after the reduction, whereas GO before the reduction is highly disordered to have small I(D)/I(G) to start with.[16, 17] Second, the increase of I(2D)/I(G) is observed and this is another indication of graphitization. SEM images of RGO/PS composites are shown in Fig. 1 (c) and (d). The inset of Fig. 1 (d) shows the typical RGO/PS composite after hot-pressing. We find RGO flakes are randomly mixed with PS. Owing to a large surface to volume ratio of RGO, RGO is shown to cover large areas of composites and the incorporation of higher concentrations of RGO reveals more crumpled or folded structures as observed previously.[13]

Fig. 2 (a) shows the electrical conductivity of the composites as a function of RGO concentration. The electrical conductivity of RGO/PS composites dramatically increases by five orders of magnitude upon increasing the RGO concentration from 0.25 vol.% to 2.5 vol.% and it starts to saturate above 2.5 vol.%. It can be explained that the electrical conductivity of the composites increases significantly once the conductive networks of RGO have been formed above a certain critical concentration of RGO in the matrix.

The trend exhibits a power law dependence on the RGO concentration. It indicates that the electrical conductivity of the composites follows the percolation model, $\sigma = \sigma_f[(\varphi - \varphi_c)/(1 - \varphi_c)]^t$, where $\sigma_f$ is the filler (RGO) conductivity, $\varphi$ is the volume loading fraction of RGO, $\varphi_c$ is the threshold volume loading fraction, and t is the critical exponent which is believed to be universal and close to 2.[18] The inset shows that the best fitting of data yields a critical exponent $t \sim 1.6$. The percolation threshold ($\varphi_c$) is ~ 0.25 vol.% and the extracted electrical conductivity of the filler ($\sigma_f$) is $1.9 \times 10^3$ S/m. The $\varphi_c$ is slightly higher than the previous reported value (~ 0.1 vol.%) in ref 13. We speculate that this may be attributed to non-uniform dispersion of RGO or our use of insulating media (PS) with a larger molecular weight. These may result in different dispersion behavior and difference in the percolation threshold. In addition, the $\sigma_f$ is two orders of magnitude lower than that in ref. 13. This may originate from imperfect reduction process.

Temperature-dependent electrical conductivity (σ) of RGO/PS composites was investigated at the temperature ranging from 30 to 300 K. The temperature-dependent σ displays an insulating behavior over the measurement temperature range as shown in Fig. 2 (b). The electrical conductivity shows the exponential dependence on temperature ($\sigma \sim \exp(-T^{-1/2})$). It can be explained by the variable-range hopping (VRH) mechanism, which deals with electrical conductivity as a function of temperature for disordered systems. The generalized form of electrical conductivity based on VRH is described by[19-21]

$$\sigma = \sigma_0 \exp\left( -\left[ \frac{T_0}{T} \right]^p \right)$$

where $\sigma_0$, $T_0$, and p are the pre-factor, the characteristic temperature, and the characteristic exponent, respectively.

There have been many reports on VRH behaviors in electrical conductivity for single-layer RGO flakes.[19, 22-24] In general, different types of VRH can be observed in 2D RGO or defective graphene systems. For example, it has been reported that graphene-derived 2D single flakes follow either Mott VRH (p = 1/3 for 2D systems)[22-24] or Efros-Shklovskii (ES) VRH (p = 1/2).[19, 25] Coulomb interactions in disordered systems can decrease the density of states near the Fermi level and this leads to p = 1/2 and the ES VRH conduction behavior.[19, 20] In contrast, p = 1/(D+1) (where D is the dimension of systems) for Mott VRH is attributed to the finite and constant density of states near the Fermi level.[19, 21] Recently, D. Joung *et al.* reported relationship between $sp^2$ domain fractions of individual RGO flakes and the ES VRH behavior. They demonstrated the ES VRH behavior (p ~ 1/2) in RGO. With increasing $sp^2$ domains, the characteristic temperature $T_0$ was reduced and the 2D localization length extracted from $T_0$ increased.[19] It indicates the imperfect reduction deteriorates the electrical conductivity due to disorder. In addition, C. Chuang *et al.* experimentally demonstrated that hydrogenated graphene could also exhibit ES VRH.[25]

We observe that the electrical conductivity of RGO/PS composites exhibits the ES VRH

behavior (p ~ 1/2, thus similar to the ES VRH behavior previously reported on individual RGO flakes[19]) as shown in Fig. 2 (b). The exponent p was also confirmed using the logarithmic derivative analysis as used in refs. 19 and 25. On the other hand, the previous studies on nano graphite composites showed 2D (p = 1/3) or 3D (p = 1/4) Mott VRH.[26, 27] We conjecture those different observations of p in composites based on graphene or nano graphite may have to do with the more conductive filler (graphene/graphite, which individually do not exhibit VRH conduction) as well as differences in degree of disorder and arrangement of fillers. The characteristic temperature shown in the inset of Fig. 2 (b) largely decreases with increasing RGO concentration, possibly related to increase of localization length.[19]

Thermal conductivity (κ) of RGO/PS composites was measured as functions of RGO concentration and temperature as presented in Fig. 3 (a) and (b). The thermal conductivity of RGO-filled composites monotonically increases as the volume concentration of RGO is increased. The thermal conductivity of the pure PS sample is measured to be ~ 0.16 W/mK, which is close to the published values (0.17 ~ 0.18 W/mK).[28, 29] The thermal conductivity of 10 vol.% RGO/PS composite is ~ 0.3 W/mK and enhanced by ~ 90 % compared with pure PS. The thermal conductivity of 20 vol.% RGO/PS composites was not characterized since the relatively rough surface made it difficult for us to fabricate the devices and do a reliable measurement.

The thermal conductivity of RGO/PS composites is comparable to the previous result of nano graphite/PS composites at around 10 vol.% (~ 0.4 W/mK).[28] However, it is lower than that of graphite/epoxy composites (~ 2.7 W/mK) or graphene/epoxy composites (~ 5.1 W/mK) at the same loading percent.[30, 31] This difference may be related to factors such as RGO with defective graphitic domains leading to the poor thermal transport properties (0.14 ~ 2.87 W/mK),[5] relatively weaker thermal linkages between polystyrene and RGO, higher thermal conductivity of graphite (80 ~ 2000 W/mK) or graphene (2000 ~ 4000W/mK),[32, 33] and slightly higher thermal conductivity of epoxy matrix (~ 0.2 W/mK).[30, 31]

In addition, the thermal conductivity of composites was characterized at temperature (T) ranging from 100 to 300 K as shown in Fig. 3 (b). The thermal conductivity of RGO/PS composites decreases almost linearly with decreasing T from 300 to 150 K and the T-dependence weakens below 150 K. From 150 to 300 K, $d\kappa/dT$ of 10 vol.% RGO composite is found to be ~ $6.9\times10^{-4}$ W/mK$^2$ and that of 5 vol.% is lower to be ~ $4.9\times10^{-4}$ W/mK$^2$. Pure PS samples do not exhibit appreciable temperature dependence over the entire measurement temperature range.

In summary, RGO/PS composites were prepared by the modified Hummers method and the chemical reduction method. The electrical conductivity of composites is significantly improved as we incorporate higher concentrations of RGO into PS. It is governed by the percolation model with the threshold of ~ 0.25 vol.% and it reaches 135 S/m at 20 vol.% RGO. We observe the electrical conductivity as a function of temperature shows the ES VRH behavior, which has not been observed in RGO composites so far. The thermal conductivity of the composites is enhanced by ~ 90 % compared with pure PS as the filler concentration is increased from 0 to 10 vol.%. However, it is not comparable to that of graphene/epoxy composites at the same loading percent.[31] On the other hand, with continued improvement of RGO reduction processes[34] and better linkage between RGO/polymer, the thermal conductivity of RGO/polymer composites has potential to be improved further, making RGO/polymer composites such as RGO/PS or RGO/epoxy a good alternative for many thermal management applications in the future.

**Acknowledgement**

We thank Dr. Dmitriy Dikin at Northwestern University for useful discussions regarding the composite preparation and thank Dr. Chin-Teh Sun at Purdue University for access to the hot-press machine. This work was partly supported by the Purdue Cooling Technologies Research Center (CTRC), a National Science Foundation (NSF) industry/university cooperative research center and by Defense Threat Reduction Agency (DTRA).

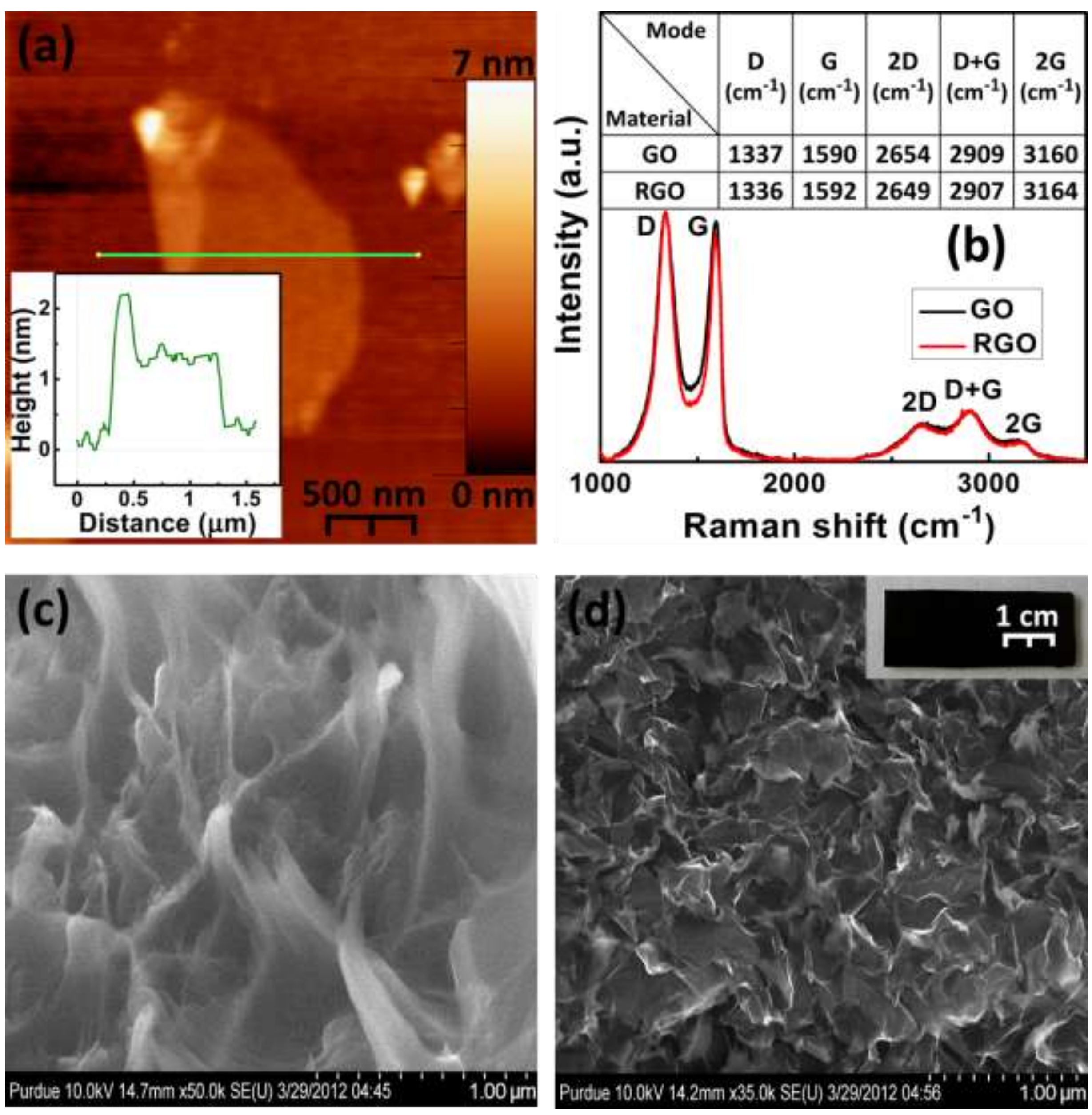

| Mode / Material | D ($cm^{-1}$) | G ($cm^{-1}$) | 2D ($cm^{-1}$) | D+G ($cm^{-1}$) | 2G ($cm^{-1}$) |
|---|---|---|---|---|---|
| GO | 1337 | 1590 | 2654 | 2909 | 3160 |
| RGO | 1336 | 1592 | 2649 | 2907 | 3164 |

FIG. 1. (a) AFM image of a GO flake (inset shows the height profile along the line). (b) Raman spectra of GO and RGO flakes. (c, d) SEM image of (c) 0.5 vol.% RGO/PS composites and (d) 5 vol.% RGO/PS composites (inset, image of a typical RGO/PS composite after hot-pressing).

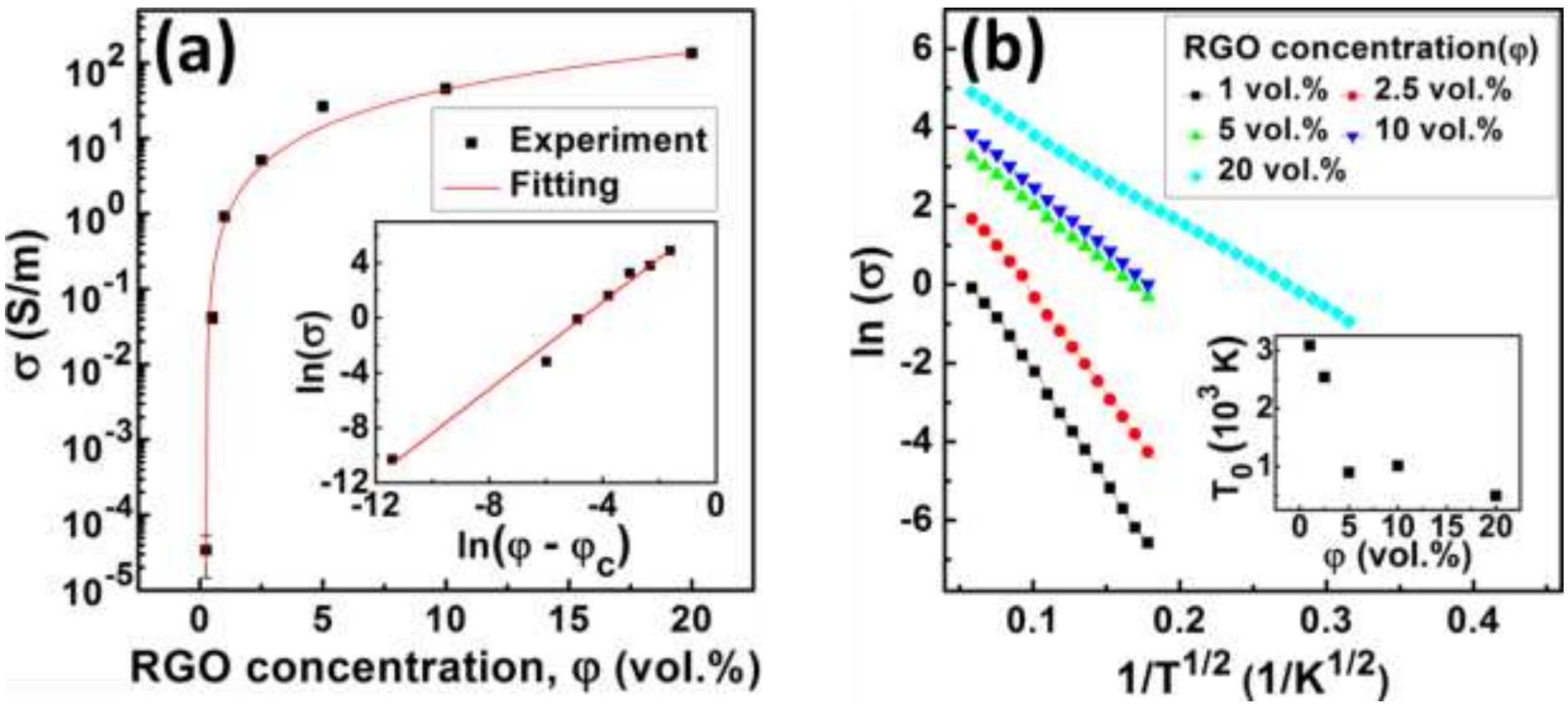

FIG. 2. (a) Electrical conductivity (σ) of RGO/PS composites with different RGO concentrations (φ) at 300 K (inset, ln(σ) vs. ln(φ - $\varphi_c$)). (b) Temperature-dependent electrical conductivity (ln(σ) vs. $1/T^{1/2}$) at various RGO concentrations (inset, $T_0$ vs. φ).

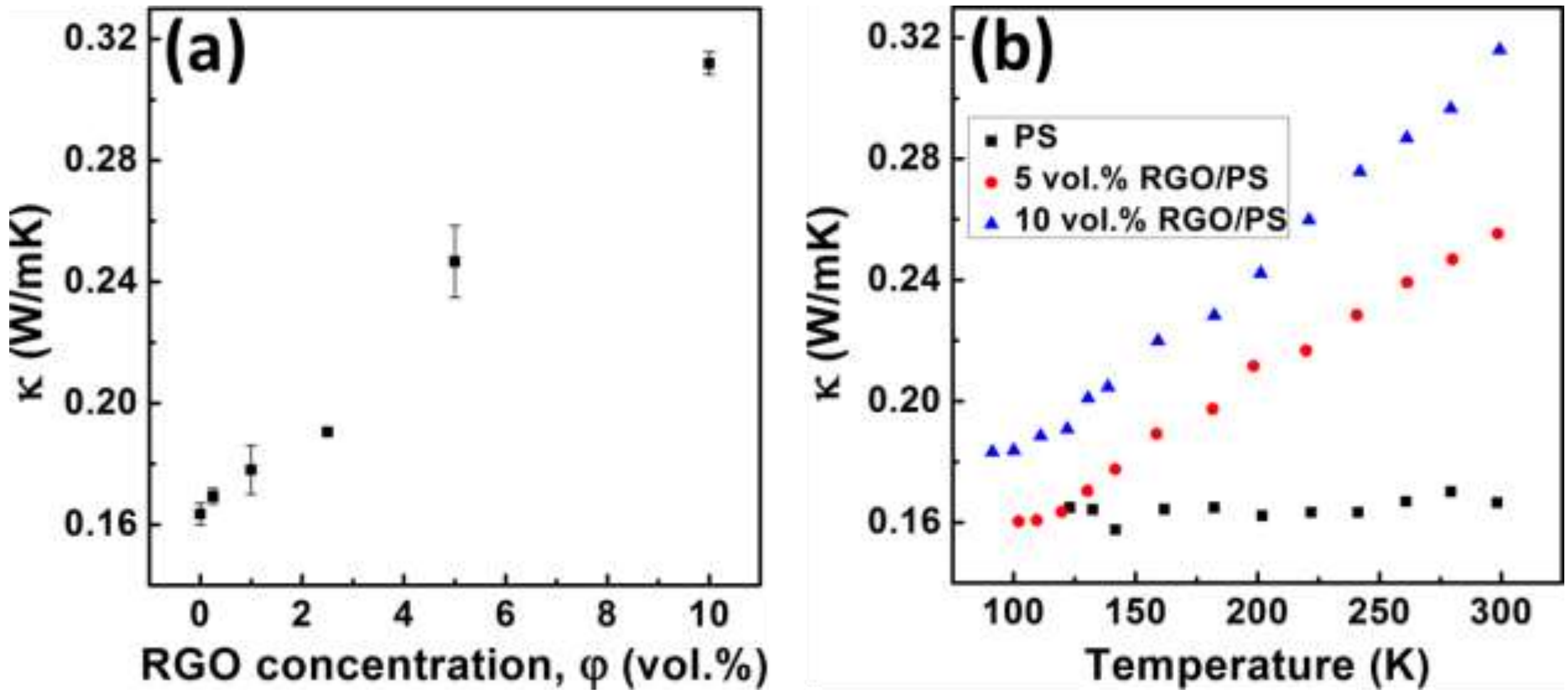

FIG. 3. (a) Thermal conductivity (κ) of RGO/PS composites with different RGO concentrations (φ) at 300 K. (b) Temperature-dependent thermal conductivity (κ) at various RGO concentrations.